\documentclass[%
reprint,
superscriptaddress,
nofootinbib,
 amsmath,amssymb,
 aps
]{revtex4-1}

\usepackage{graphicx}
\usepackage{dcolumn}
\usepackage{bm}
\usepackage{slashed}
\usepackage[colorlinks=true, pdfstartview=FitV, linkcolor=red, citecolor=blue, urlcolor=blue]{hyperref}
\usepackage{slashed}
\usepackage{array}
\usepackage[caption=false]{subfig}

\graphicspath{{./figures/}}

\newcommand{\im}{\mathrm{i}}
\newcommand{\diff}{{\mathrm{d}}}
\newcommand{\mL}{\mathrm{L}}
\newcommand{\mR}{\mathrm{R}}
\newcommand{\mF}{\mathrm{F}}
\newcommand{\bs}{\boldsymbol}

\begin{document}

\preprint{XXX}

\title{Giant photocurrent in asymmetric Weyl semimetals from the helical magnetic effect}

\author{Dmitri E. Kharzeev}
\email[]{dmitri.kharzeev@stonybrook.edu}
\affiliation{Department of Physics and Astronomy, Stony Brook University, Stony Brook, New York 11794-3800, USA}
\affiliation{Department of Physics, Brookhaven National Laboratory, Upton, New York 11973-5000}
\affiliation{RIKEN-BNL Research Center, Brookhaven National Laboratory, Upton, New York 11973-5000}

\author{Yuta Kikuchi}
\email[]{yuta.kikuchi@riken.jp}
\affiliation{RIKEN-BNL Research Center, Brookhaven National Laboratory, Upton, New York 11973-5000}

\author{Ren\'e Meyer}
\email[]{rene.meyer@physik.uni-wuerzburg.de}%
\affiliation{Institute for Theoretical Physics and Astrophysics, University of W\"urzburg, 97074 W\"urzburg, Germany}

\author{Yuya Tanizaki}
\email[]{yuya.tanizaki@riken.jp}
\affiliation{RIKEN-BNL Research Center, Brookhaven National Laboratory, Upton, New York 11973-5000}

\date{\today}

\begin{abstract}
We propose a new type of photoresponse induced  in asymmetric Weyl semimetals in an external magnetic field. In usual symmetric Weyl semimetals in a magnetic field, 
the particles and holes produced by an incident light in different Weyl cones have opposite helicities and hence move in opposite directions, canceling each others's contributions to the photocurrent. However this cancelation does not occur if the Weyl semimetal possesses both a broken particle-hole symmetry and a broken spatial inversion symmetry. We call the resulting generation of photocurrent the {\it helical magnetic effect} because it is induced by the helicity imbalance in a magnetic field. We find that due to the large density of states in a magnetic field, the helical magnetic effect induces a remarkable large photocurrent for incident THz frequency light. This suggests a potential application of asymmetric Weyl semimetals for creating THz photosensors.

\end{abstract}

\maketitle




\section{\label{sec:1}
Introduction}

The hallmark of Dirac and Weyl semimetals (DSMs/WSMs) is the emergence of massless fermion quasiparticles near the band-touching points \cite{Klinkhamer:2004hg,Volovik:2003fe,Wan:2011udc}. These low-energy excitations are described by the Dirac and Weyl equations and allow to study the effects of relativistic quantum field theory in condensed matter systems. These effects include the Chiral Magnetic Effect (CME) \cite{Fukushima:2008xe,Son:2012wh,Son:2012bg,Zyuzin:2012tv,Basar:2013iaa,vazifeh2013electromagnetic,goswami2013axionic} -- the CP-odd dissipationless transport phenomenon stemming from the chiral anomaly \cite{Adler:1969gk,Bell:1969ts,Nielsen:1983rb}. The CME has been observed in experiments on magnetotransport in a growing number of DSMs \cite{Li:2014bha,kim2013dirac,xiong2015evidence,li2015giant} and WSMs \cite{huang2015observation,wang2015helicity,zhang2015observation,yang2015observation,shekhar2015large,yang2015chiral}. 

Recently it has been pointed out that in condensed matter systems (where the Lorentz and rotational symmetries are in general absent), new types of DMSs/WSMs possessing exotic dispersion relations around the band-touching points could emerge, such as type-II WSMs \cite{Soluyanov:2015cn,Yong:2015}, multi-WSMs \cite{fang2012multi}, and asymmetric WSMs \cite{kharzeev2016chiral}. In particular, the asymmetric WSMs (aWSMs), with different dispersion relations for the left- and right-handed chiral fermions, are expected to provide a variety of new transport phenomena which are absent in ordinary relativistic systems \cite{ishizuka2016emergent,chan2016photocurrents,kharzeev2016chiral,Mukhergee1:2018,Mukhergee2:2018,Mukhergee3:2018}.

The photovoltaic effect, normally realized in semiconductors, is hard to realize in DMSs/WSMs due to severe cancellations of the induced photocurrent resulting from the Lorentz or rotational symmetries of the quasiparticle spectrum. In aWSMs, however, it is possible to avoid such cancellations -- this has been  recently proposed in \cite{chan2016photocurrents}\footnote{In Ref.~\cite{kharzeev2016chiral}, we discussed the AC CME in Weyl semimetals that break  both spatial inversion and reflection symmetry. For the photovoltaic effect, however, reflection symmetry does not need to be broken by the band structure of the material since the external light breaks it as well, as pointed out in Ref.~\cite{chan2016photocurrents}.}. The photoresponse in DMSs/WSMs for far-infrared incident light is potentially strong due to the absence of the energy  gap in the spectrum of quasiparticles.

\begin{figure}
 \begin{center}
 \includegraphics[width=7.5cm]{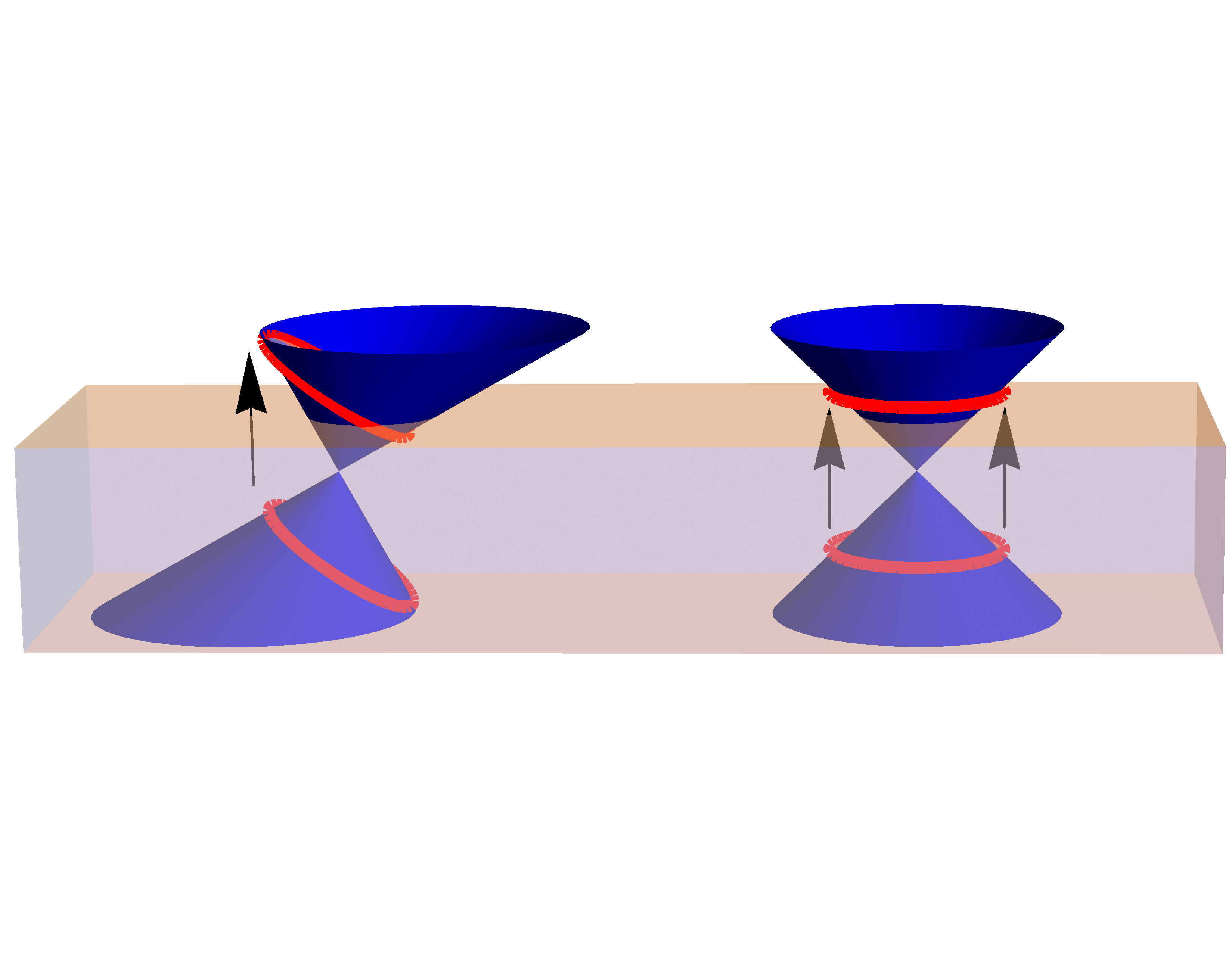}
 \vspace{-1.5cm}
 \caption{(Color online) Left and right-handed tilted Weyl cones in aWSM with the Fermi sea represented by semitransparent region. The horizontal plane parametrizes the momentum space $(k_x,k_y)$ with fixed $k_z$ and the vertical axis parametrizes the energy $E(\bs{k})$. For a fixed frequency of external light, the states below Fermi surface (the lower red line) can be excited to above the surface (the upper red line) as indicated by the black arrows at zero temperature. The vertical distance between two red lines corresponds to the excitation energy provided by the external light.
Since a part of the upper red line is below the Fermi surface in the tilted cone the excitation is highly suppressed by the Pauli blocking at low temperature.}
 \label{fig:cones}
 \end{center}
\end{figure}

In this letter, we show that in aWSMs a new type of photocurrent is induced along the direction of an external magnetic field. We will find that the corresponding photocurrent is enhanced by the density of fermion states in magnetic field, and appears very strong for the THz frequency light.
Let us begin by considering two cancellation mechanisms of photoresponse in WSMs under external magnetic field and the ways of circumventing them.
A seed of the response is provided by the particle-hole excitations driven by the incident light. The particle and hole, however, have opposite helicities, and hence move  parallel and antiparallel to the external magnetic field, which results in vanishing net current. To avoid this cancellation, we need to generate imbalance between particle and hole excitations, i.e., broken particle-hole symmetry (see Appendix~\ref{app:1} for a discussion of discrete symmetries in Weyl semimetals).
The other problem is the cancellation between two Weyl cones with opposite chiralities. Even if particle-hole asymmetry allows a finite current from the left- and right-handed Weyl cones respectively, they would cancel in inversion-symmetric systems.

Therefore, a necessary condition for the finite current to be induced is the existence of Weyl cones with broken particle-hole symmetry as well as the broken spatial inversion symmetry. Here, the particle-hole transformation acts on the energy dispersion of a single Weyl cone as $E_{\bs{k}}\to -E_{\bs{k}}+2E_0$, where $E=E_{\bs{k}}$ is dispersion relation of a Weyl cone and the Weyl point sits on $E_0$. The spatial inversion exchanges the left and right handed Weyl cones. 
One of the crucial ingredients is the asymmetry in particle-hole excitation, which effectively generates a  helicity imbalance under magnetic field. Hence, we call the photoresponse mechanism the {\it helical magnetic effect}.
\footnote{The anomalous transport of neutrino induced by helical imbalance, called helical magnetic effect, was discussed in the context of supernovae in~\cite{Yamamoto:2015gzz}}
A typical dispersion relation for a pair of Weyl cones which induces the effect is shown in Fig.~\ref{fig:cones}. Different dispersion relations for left- and right-handed Weyl cones are allowed due to the broken spatial inversion symmetry. Furthermore, the cone on the left is tilted, which breaks the particle-hole symmetry. Therefore, as we will see explicitly, aWSMs possessing a pair of inversion and particle-hole symmetry-breaking Weyl cones will exhibit the helical magnetic effect. 

We will find that the helical magnetic effect induces a significantly larger photocurrent in IR region compared to the photocurrent discussed in \cite{chan2016photocurrents}. We show by explicit calculation based on chiral kinetic theory that the current is particularly large if the incident light is of  teraherz frequency. The same effect was discussed in \cite{Golub2017} by considering the Landau quantization under strong magnetic field, that is complementary to our chiral kinetic approach as it is valid under weak magnetic field. This result points out the possibility of realizing very efficient infrared radiation sensors in aWSMs such as TaAs, WTe$_2$ \cite{Soluyanov:2015cn}, or SrSi$_2$ \cite{huang2016new}.



\section{\label{sec:2}
Photoresponse in Weyl semimetals}

In this section we derive general formulae for the photovoltaic effect in WSMs   \cite{chan2016photocurrents} taking into account the effect of external magnetic field.
Consider the Weyl hamiltonian in the Brillouin zone $\bs{q}=(q_x,q_y,q_z)$ described by the following two band model around a left-handed Weyl point,
\begin{align}
 \label{eq:Weyl_ham}
 H_{\mL}(\bs{q})=\hbar v_\mF\bs{q}\cdot\bs{\sigma} + \hbar \bs{v}_{\mL}\cdot\bs{q}.
\end{align}
$v_\mF$ is the Fermi velocity and $\sigma_i$ is the Pauli matrix. The second term tilts the Weyl cone in the direction determined by a constant vector $\bs{v}_{\mL}$.
The interaction between the emergent Weyl fermions and an incident light $\bs{A}(t)=\bs{A}_{+}e^{\im\omega t}+\bs{A}_{-}e^{-\im\omega t}$ with frequency $\omega$ is introduced by the Peierls substitution $H(\bs{q})\rightarrow H(\bs{q} - \frac{e}{\hbar}\bs{A})$. 
The interaction hamiltonian $V(t)$ becomes
\begin{align}
 V(t) = V_{+}e^{\im\omega t}+V_{-}e^{-\im\omega t},
\end{align}
with $V_{\pm} = - e v_{\mF}\sum_{i=x,y,x}A_{\pm,i}\sigma_i$. It is noted that, although the Peierls substitution in the second term in Eq.~\eqref{eq:Weyl_ham} also yields an interaction, it does not contribute to the transition probability because it is diagonal in spin space.

We compute the electric current induced by the electron-photon interaction under the influence of a static background magnetic field $\bs{B}$,
\begin{align}
 \label{eq:current}
 &\bs{J}=\bs{J}_\mathrm{photo}+\bs{J}_\mathrm{HME},
 \\
 \label{eq:photocurrent}
 &\bs{J}_\mathrm{photo}=-\frac{e}{\hbar}\sum_{l=\pm}\int\frac{\diff^3q}{(2\pi)^3}
 \frac{\partial E_l}{\partial\bs{q}}[f_l(\bs{q})-f_l^0(\bs{q})],
 \\
 \label{eq:HMcurrent}
  &\bs{J}_\mathrm{HME} =\frac{e^2}{\hbar^2}\bs{B}\sum_{l=\pm}\int\frac{\diff^3q}{(2\pi)^3}
 \left(\frac{\partial E_l}{\partial\bs{q}}\cdot\bs{\Omega}_l\right)[f_l(\bs{q})-f_l^0(\bs{q})],
\end{align}
where $l=\pm$ label upper and lower part of a single Weyl cone and $E_{\pm}$ are the corresponding dispersion relations obtained from the Weyl Hamiltonian~\eqref{eq:Weyl_ham}. 
Eq.~\eqref{eq:photocurrent} is the photocurrent discussed in \cite{chan2016photocurrents}. Chiral kinetic theory gives the expression Eq.~\eqref{eq:HMcurrent} under the external magnetic field and Berry curvature \cite{Son:2012wh,Stephanov:2012ki,Chen:2012ca}, that corresponds to the current induced by the helical magnetic effect.
The Berry curvature $\bs{\Omega}_l$ is given by
\begin{align}
 \Omega^i_{\pm} \equiv -\im\epsilon^{ijk}\frac{\partial u_\pm^\dagger}{\partial q^j}\frac{\partial u_\pm}{\partial q^k} = -\Omega^i_{\mp} ,
\end{align}
with $u_\pm$ being the eigenstates of Weyl hamiltonian \eqref{eq:Weyl_ham}, 
$f_\pm$ and $f_\pm^0$ denote nonequilibrium and equilibrium distribution functions, and each sign again corresponds to the upper and lower Weyl cone, respectively.
The $f_\pm(\bs{q})$ are obtained by solving the following equations,
\begin{align}
 \label{eq:n+}
 \frac{\diff f_+}{\diff t} 
 &= \Gamma_{\{-\rightarrow+\}}f_-(1-f_+) -\Gamma_{\{+\rightarrow-\}}f_+(1-f_-) 
 \nonumber\\
 &- \frac{f_+-f^0_+}{\tau},
 \\
 \label{eq:n-}
 \frac{\diff f_-}{\diff t} 
 &= -\Gamma_{\{-\rightarrow+\}}f_-(1-f_+) +\Gamma_{\{+\rightarrow-\}}f_+(1-f_-) 
 \nonumber\\
 &- \frac{f_--f^0_-}{\tau},
\end{align}
where the second terms in the right hand side of both equations are relaxation terms induced by the interaction between excited electron and the medium in the relaxation time approximation with a relaxation time $\tau$. The rates $\Gamma_{\{\mp\rightarrow\pm\}}$ express absorption and emission processes, respectively,  given by Fermi's golden rule,
\begin{align}
 \Gamma_{\{-\rightarrow+\}} &= \frac{2\pi}{\hbar}|\left<q_+|V_+|q_-\right>|^2 \delta(\Delta E-\hbar\omega)
 \nonumber\\
 &=\Gamma_{\{+\rightarrow-\}}
 \equiv \Gamma\,,
\end{align}
where $\Delta E\equiv E_+-E_-$ is electron-hole excitation energy.
By solving Eqs.~\eqref{eq:n+} and \eqref{eq:n-} and assuming $\tau\Gamma\ll1$ we obtain the distribution functions
\begin{align}
 f_{+} &=f^0_{+}+\tau\Gamma\left[f^0_--f^0_+\right],
 \\
 f_{-} &=f^0_{-}-\tau\Gamma\left[f^0_--f^0_+\right],
\end{align}
which give the stationary solutions of Eqs.~\eqref{eq:n+} and \eqref{eq:n-}, respectively.
Substituting these distribution functions into Eq.~\eqref{eq:current} and summing up the electron and hole contributions, we obtain
\begin{align}
 \label{eq:PVEcurrent}
 \bs{J}_\mathrm{photo}
 &= -\frac{e}{\hbar}\int\frac{\diff^3q}{(2\pi)^3}
 \frac{\partial (E_+-E_-)}{\partial\bs{q}}
 \tau\Gamma[f^0_- - f^0_+]
 \nonumber\\
 &=\left(\frac{-e^3 \tau\omega^2A^2}{16 \pi^2 \hbar^2}\right)\bar{\bs{J}}_\mathrm{photo}(\omega),
 \\
 \label{eq:HMEcurrent}
 \bs{J}_\mathrm{HME}
 &=\frac{e^2}{\hbar^2}\bs{B}\int\frac{\diff^3q}{(2\pi)^3}
 \left(\frac{\partial (E_++E_-)}{\partial\bs{q}}\cdot\bs{\Omega}_+\right)
 \tau\Gamma[f^0_- - f^0_+]
 \nonumber\\
 &=\left(\frac{e^4 \tau v_L v_FA^2B}{16 \pi^2 \hbar^3}\right)\bar{\bs{J}}_\mathrm{HME}(\omega),
\end{align}
where the dimensionless currents $\bar{\bs{J}}_\mathrm{photo}$ and $\bar{\bs{J}}_\mathrm{HME}$ are given by
\begin{align}
 \label{eq:dimlesscurrent}
 \bar{\bs{J}}_\mathrm{photo}(\omega)
 &=4\int \diff^3\left(\frac{v_Fq}{\omega}\right)
 \frac{\partial (E_+-E_-)}{\partial\hbar v_F\bs{q}}
 \nonumber\\
 &\times\left|\left<q_+\left|\frac{V_+}{(- e v_F A)}\right|q_-\right>\right|^2 
 \nonumber\\
 &\times\delta\left(\frac{\Delta E}{\hbar\omega}-1\right)[f^0_- -f^0_+]
 \\
 \label{eq:dimlessHME}
 \bar{\bs{J}}_\mathrm{HME}(\omega)
 &=4\frac{v_F}{v_L}\bs{\hat{B}}\int \diff^3\left(\frac{v_Fq}{\omega}\right)
 \left(\frac{\partial (E_++E_-)}{\partial\hbar v_F\bs{q}}\cdot\frac{\omega^2\bs{\Omega}_+}{v_F^2}\right)
 \nonumber\\
 &\times\left|\left<q_+\left|\frac{V_+}{(- e v_F A)}\right|q_-\right>\right|^2 
 \nonumber\\
 &\times\delta\left(\frac{\Delta E}{\hbar\omega}-1\right)[f^0_- -f^0_+].
\end{align}
We remark on the broken symmetries necessary for the helical magnetic effect:
\begin{itemize}
\item $\bs{J}_\mathrm{HME}$ vanishes if each Weyl cone is particle-hole symmetric, i.e., $E_++E_-=0$, while $\bs{J}_\mathrm{photo}$ yields a finite contribution.
\item The sum of left and right handed contributions to $\bs{J}_\mathrm{HME}$ yields a finite total current only if spatial inversion symmetry is broken.
\end{itemize}
The various properties of Eqs.~\eqref{eq:PVEcurrent} and \eqref{eq:dimlesscurrent} were discussed in \cite{chan2016photocurrents}.


\section{\label{sec:helical}
Helical Magnetic Effect}

We consider the photoresponse induced by linearly polarized light in an external magnetic field, i.e., the helical magnetic effect, given by $\bs{J}_\mathrm{HME}$ in Eq.~\eqref{eq:HMcurrent}.
The current is induced when both particle-hole symmetry and spatial inversion symmetry are absent. 
The helical magnetic effect is similar to chiral magnetic effect, which induces the dissipationless current in the direction of external magnetic field when particle-hole symmetry of the left and right Weyl cones are separately broken by a finite chemical potential and spatial inversion symmetry is broken by a chiral chemical potential.
The helical magnetic effect is, however, different in that it is induced by incident light without the need for a chiral chemical potential.

Now, we take a closer look at Eq.~\eqref{eq:HMcurrent},
\begin{align}
\label{eq:helical2}
   \bs{J}_\mathrm{HME} 
 &=\frac{e^2}{\hbar^2}\bs{B}\int\frac{\diff^3q}{(2\pi)^3}
 \left(\frac{\partial (E_++E_-)}{\partial\bs{q}}\cdot\bs{\Omega}_+\right)
 \nonumber\\
 &\times\tau\Gamma[f^0_-(\bs{q}) - f^0_+(\bs{q})].
\end{align}
The broken particle-hole symmetry makes the terms inside the bracket finite and the cancellation of photocurrent inside a single Weyl cone can be avoided.
We calculate the current for a left-handed tilted Weyl cone \footnote{Other types of WSMs with broken particle-hole symmetry may be realized by replacing the second term in the following hamiltonian.},
\begin{align}
\label{eq:tilted_Weyl}
 H_\mL = \hbar v_\mF \bs{q}\cdot\bs{\sigma} + \hbar v_\mL q_z.
\end{align}
which yields two eigenenergies, $E_\pm =\pm \hbar v_{\mF}q+\hbar v_{\mL}q_z$, with corresponding eigenvectors,
\begin{align}
 \label{eq:eigenvector}
 \left|q_{\pm}\right>=&\frac{1}{\sqrt{2q(q\mp q_z)}}
 \begin{pmatrix}
  q_x-\im q_y \\
  \pm q-q_z
 \end{pmatrix},
\end{align}
where $q\equiv|\bs{q}|$.
$\partial(E_++E_-)/\partial\bs{q}=2\hbar v_\mL\hat{q}_z$ in Eq.~\eqref{eq:helical2} is indeed nonvanishing due to the second term of \eqref{eq:tilted_Weyl}, which explicitly breaks particle-hole symmetry.

The interaction between the electrons and linearly polarized light is given by $V_i=- e v_\mF A\sigma_i$, where $i=x,y,z$ is a direction of the polarization. Although the second term in Eq.~\eqref{eq:tilted_Weyl} also induces the interaction between the incident light and fermions we ignore it as it does not contribute to the transition probability \eqref{eq:transition_linear}. The corresponding transition probability is given by
\begin{align}
 \label{eq:transition_linear}
 \left|\left<q_+\left|\frac{V_i}{(-e v_{\mF}A)}\right|q_-\right>\right|^2
 =\frac{q^2-q_i^2}{q^2}.
\end{align} 

Assuming incident light polarized in $z$ direction (setting $i=z$ above), we can simplify the dimensionless helical magnetic current to
\begin{align}
\label{eq:left_current}
 \bar{\bs{J}}^\mL_\mathrm{HME}
 &=4\frac{v_\mF}{v_\mL}\bs{\hat{B}}\int \diff^3\left(\frac{v_{\mF}q}{\omega}\right)
 \frac{\partial (E_++E_-)/\hbar}{\partial v_\mF\bs{q}}\cdot\frac{\hat{\bs{q}}}{2v_{\mF}^2q^2/\omega^2}
 \nonumber\\
 &\times\frac{q^2-q_z^2}{q^2}\delta\left(\frac{\Delta E}{\hbar\omega}-1\right)[f^0_-(\bs{q})-f^0_+(\bs{q})]
 \nonumber\\
 &=4\pi\bs{\hat{B}}\int d\cos\theta\cos\theta(1-\cos^2\theta)
 \nonumber\\
 &\times\big[f^0[((v_{\mL}/v_{\mF})\cos\theta-1)\hbar\omega/2]
 \nonumber\\
 &-f^0[((v_{\mL}/v_{\mF})\cos\theta+1)\hbar\omega/2]\big],
\end{align}
where the equilibrium distribution functions are given by $f^0_\pm\equiv f^0(E_\pm)\equiv[\mathrm{e}^{(E_{\pm}-\mu)/T}+1]^{-1}$.
The contribution from a right-handed Weyl cone with a hamiltonian 
\begin{align}
H_\mR =-\hbar v_\mF \bs{q}\cdot\bs{\sigma} - \hbar v_\mR q_z
\end{align}
is similarly calculated to be
\begin{align}
 \bar{\bs{J}}^\mR_\mathrm{HME}
 &=-4\pi\bs{\hat{B}}\int \diff\cos\theta\cos\theta(1-\cos^2\theta)
 \nonumber\\
 &\times\big[f^0[((v_\mR/v_\mF)\cos\theta-1)\hbar\omega/2]
 \nonumber\\
 &-f^0[((v_\mR/v_\mF)\cos\theta+1)\hbar\omega/2]\big],
\end{align}
where we used $\bs{\Omega^+}=-\bs{\Omega^-}$. In WSMs with spatial inversion symmetry, $v_L=v_R$ and hence $\bs{J}^\mL_\mathrm{HME}=-\bs{J}^\mR_\mathrm{HME}$ which leads to cancellation between two nodes.
Therefore we need an inversion-breaking asymmetry between left and right-handed Weyl cones. For example, if we take $v_\mL\neq0$ and $v_\mR=0$ as shown in Fig.~\ref{fig:cones}, the contribution from right-handed Weyl cone vanishes. The induced current is given only by the left-handed contribution \eqref{eq:left_current}. If time reversal symmetry is not broken we need another pair of Weyl cones, i.e., two left- and two right-handed Weyl cones, and two left/right-handed cones are exchanged under time reversal transformation.

The helical magnetic effect can be also generated by circularly polarized light as well, in which case the computation is completely analoguous and we do not repeat it here.


\section{\label{sec:5}
Discussion}

In this section we discuss the magnitude and frequency dependence of the helical magnetic current under realistic assumptions for the input parameters. 

Let us first estimate the magnitude of the  current induced by the helical magnetic effect in comparison with the photocurrent given by \eqref{eq:PVEcurrent} and \eqref{eq:dimlesscurrent} in the same system, with the directions of tilt and polarization are taken to be the same.
More specifically, we take the tilt of Weyl cones, polarization of incident light, and external magnetic field along the $z$ axis, and measure the currents induced by the helical magnetic effect and photovoltaic effect in $z$-direction.
Under this setup, the dimensionless parts of currents $\bar{\bs{J}}_\mathrm{photo}$ and $\bar{\bs{J}}_\mathrm{HME}$ have the same frequency dependence, $|\bar{\bs{J}}_\mathrm{photo}|(\omega)=\frac{1}{2}|\bar{\bs{J}}_\mathrm{HME}|(\omega)$.
We note that $\bs{J}_\mathrm{HME}/|\bar{\bs{J}}_\mathrm{HME}|$ is not suppressed in the IR region in contrast to $\bs{J}_\mathrm{photo}/|\bar{\bs{J}}_\mathrm{photo}|$.
The ratio of prefactor of the helical magnetic current \eqref{eq:HMEcurrent} and photocurrent \eqref{eq:photocurrent} is
\begin{align}
 \left.\left(\frac{e^4 \tau v_\mF v_\mL A^2B}{16 \pi^2 \hbar^3}\right)\right/\left(\frac{e^3 \tau\omega^2A^2}{16 \pi^2 \hbar^2}\right)
 =\frac{ev_\mF v_\mL B}{\hbar(2\pi\nu)^2}\sim 3.8\times 10^2,
\end{align}
for magnetic field 1~T, frequency of incident light $\nu=\omega/2\pi =0.1$~THz, $v_\mL/v_\mF=0.1$, and $v_\mF=c/300$ where $c$ is the speed of light. 
It is remarkable that the helical magnetic effect yields a much larger current (by a factor of a hundred) in the infrared region compared with the normal photocurrent, the latter being suppressed in the IR by factor $\omega^2$. This enhancement of the helical magnetic effect is due to the large density of fermion states in magnetic field $\sim eB$ as opposed to the usual density of states $\sim \omega^2$ that determines the magnitude of conventional photocurrent.

Next, for the purpose of estimating the frequency dependence of HME response analytically, let us further simplify Eq.~\eqref{eq:left_current} in a reasonable parameter region $v_\mL/v_\mF\ll1$,
\begin{align}
 \bar{\bs{J}}^\mL_\mathrm{HME}
 &=\frac{2\pi}{15}\frac{\hbar\omega}{T}\frac{v_\mL}{v_\mF}\hat{B}
 \nonumber\\
 &\times
 \left(\frac{1}{\cosh^2(\frac{\hbar\omega}{4T}+\frac{\mu}{2T})}-\frac{1}{\cosh^2(\frac{\hbar\omega}{4T}-\frac{\mu}{2T})}\right),
\end{align}
Hence, for a fixed intensity $I= \epsilon_0c(\omega A)^2$, the $\omega$ dependence of $\bs{J}^\mL_\mathrm{HME}$ is given by
\begin{align}
 |\bs{J}^\mL_\mathrm{HME}|
 &\propto \left(\frac{\hbar\omega}{T}\right)^{-1}
 \left(\frac{1}{\cosh^2(\frac{\hbar\omega}{4T}+\frac{\mu}{2T})}-\frac{1}{\cosh^2(\frac{\hbar\omega}{4T}-\frac{\mu}{2T})}\right).
\end{align}
At low temperature $T\ll\mu$, since the first term in the parenthesis is exponentially suppressed for $\mu>0$, the peak location is roughly estimated as $\nu_\text{peak}=\omega_\text{peak}/2\pi=\mu/\pi\hbar$ with the peak width proportional to $T/\hbar$.

\begin{figure}[t]
\centering
\begin{minipage}{.49\textwidth}
\subfloat[$T=20$ K]{
\includegraphics[width=8.5cm]{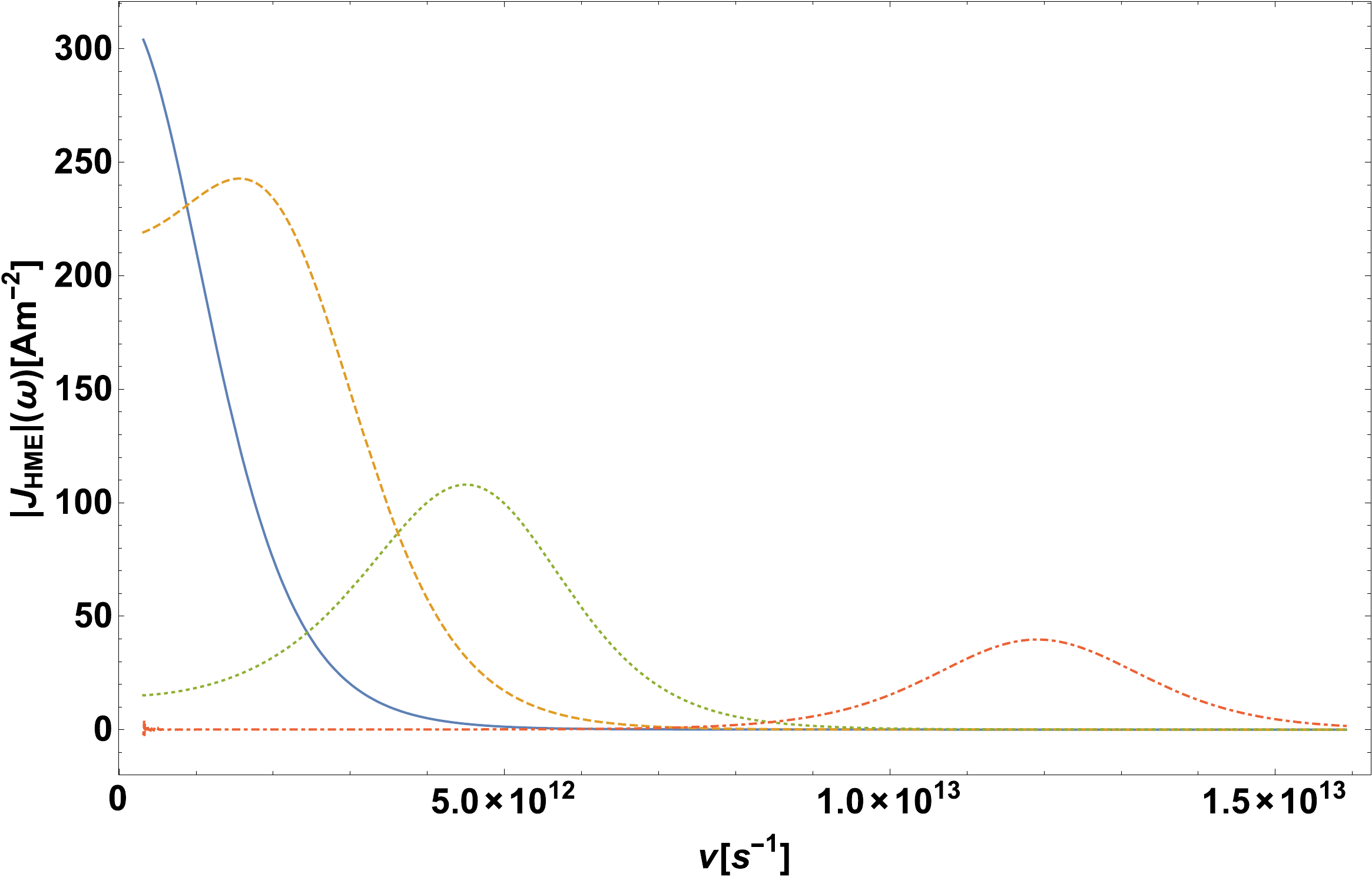}
\label{fig:5.1}
}\end{minipage}\\ \vspace{2mm}
\begin{minipage}{.49\textwidth}
\subfloat[$T=77$ K]{
\includegraphics[width=8.5cm]{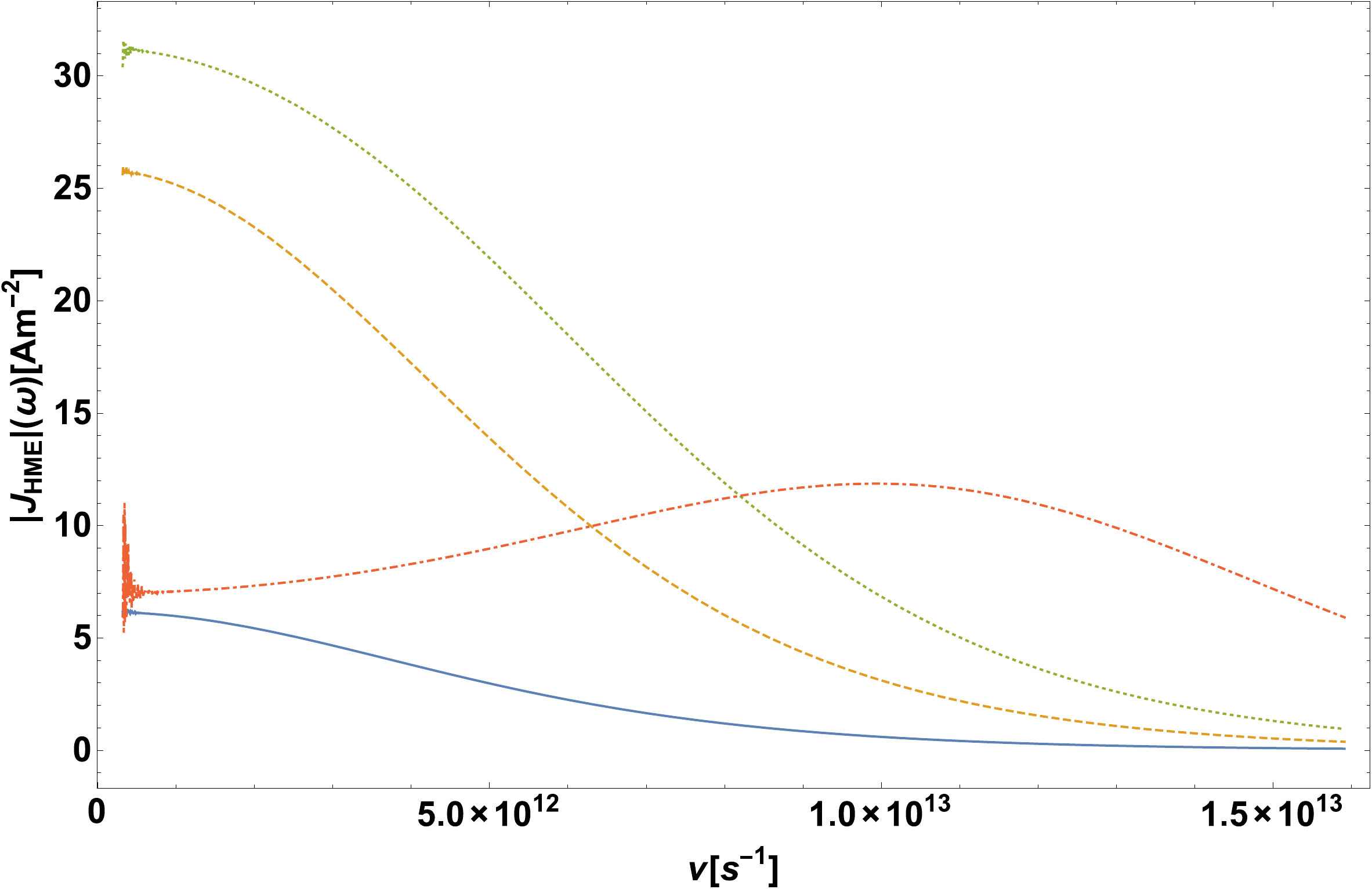}
\label{fig:5.2}
}\end{minipage}
\caption{(Color online) The helical magnetic current as a function of the frequency of external light $\nu (=\omega/2\pi)$ for chemical potentials $\mu=1,\, 5,\,10,\, 25$ meV (solid blue, dashed orange, dotted green, and dot-dashed red, respectively), at the temperature $T=20$K (a) and $T=77$K (b). The other input parameters are $v_\mF=c/300$, $v_\mL/v_\mF=0.1$, $B=1$T, $\tau = 10^{-11}$s, and laser intensity $I= \epsilon_0c(2\pi \nu A)^2=1.5\times10^6$ Wm$^{-2}$.}
\label{fig:5}
\end{figure}

Finally, we compute the magnitude of the helical magnetic current in THz regime ($1\sim 100$ THz).
The relaxation times in Weyl semimetals are typically 
10 ps, and typical laser intensity is $I= \epsilon_0c(\omega A)^2\sim10^6$ Wm$^{-2}$.  In fig.~\ref{fig:5}, we plot the helical magnetic current  \eqref{eq:left_current} as a function of the frequency of the incident light at temperatures $T=20$ K and $T=77$ K. At $T=20$ K (fig.~\ref{fig:5.1}), the peak shifts towards the infrared frequency (towards the ultraviolet frequency) for smaller (larger) chemical potential $\mu$. Furthermore, the maximal value of the current \eqref{eq:left_current} increases strongly with lowering $\mu$. 
On the other hand, at $T=77$ K (fig.~\ref{fig:5.2}), the strongest response is obtained for $\mu\sim10$ meV and the incident light with the frequency $\sim 0.1$ THz although the peak of the current is smeared compared with that at low temperature.



\section{\label{sec:6}
Conclusions}

We proposed a new type of photoresponse in aWSMs -- the helical magnetic effect.
The current is induced in the direction of an external magnetic field when the WSM breaks the spatial inversion and particle-hole symmetries.
The spatial inversion symmetry is broken in aWSMs with different dispersion relations for the left and right handed Weyl nodes. The particle-hole symmetry is typically broken in tilted WSMs including the recently proposed type-II WSMs \cite{Soluyanov:2015cn}.

The induced current was computed here in the semiclassical limit described by chiral kinetic theory. The excitation around the Fermi surface induced by the incident light combined with the Berry curvature yields a nonvanishing response current in tilted aWSMs. 
We estimated the magnitude of induced current, and found that the helical magnetic effect yields much larger currents compared to the standard photoresponse for infrared incident light. Furthermore, we analyzed the frequency dependence of the induced current for various temperatures and chemical potentials. The infrared sensitivity of the helical magnetic effect is quite sharp at $T=20$ K with a well-defined peak around the chemical potential, and still large even at $T=77$ K although the peak is strongly broadened at these higher temperatures and weakened in magnitude.

The helical magnetic effect can be realized, for instance, in WTe$_2$ which is proposed as a type-II WSM, with an expected asymmetry for left and right-handed Weyl cones. TaAs and SrSi$_2$ are other candidate aWSMs. The strong photoresponse predicted in this paper may allow to create an efficient photosensor for THz radiation using the aWSMs.

\begin{acknowledgements}
We thank L.~E.~Golub for useful comments on the manuscript.
The work of D.K. was supported by the
U.S. Department of Energy 
under contracts No. DE-FG-88ER40388, DE-SC-0017662 and DE-AC02-98CH10886.
Y.K. was supported by the Grants-in-Aid for JSPS fellows (Grant No.15J01626). Y.K. appreciates the hospitality of Department of Physics and Astronomy, Stony Brook University. The work of R.M. was supported in part by the U.S. Department of Energy under Contract No. DE-FG-88ER40388, by the Alexander-von-Humboldt Foundation through a Feodor Lynen postdoctoral fellowship, as well as by the DFG through SFB1170 "ToCoTronics". 
The work of Y.T. is supported by the RIKEN special postdoctoral program. 
\end{acknowledgements}


\appendix

\section{\label{app:1}
C, P, T transformations in Weyl semimetals}

We summarize the $C,P,T$ transformations useful for discussion of deformed Weyl cones.
The particle-hole symmetry used in the main text corresponds to $CP$-symmetry as described below.
First, we collect the transformation law of creation and annihilation operators of a four-component Dirac fermion $\Psi_{\bs{k}}$ in momentum space:
\begin{align}
 C\Psi_{\bs{k}}C^{-1}&=\im\gamma_2\Psi^*_{-\bs{k}},
 \\
 P\Psi_{\bs{k}}P^{-1}&=\gamma_0\Psi_{-\bs{k}},
 \\
 \label{eq:time_rev}
 T\Psi_{\bs{k}}T^{-1}&=\im\gamma_1\gamma_3\Psi_{-\bs{k}},
\end{align}
where $T$ is an anti-unitary operator.
Notice that $C$ and $P$ flip the chirality. We use $P$ transformation to characterize the asymmetry of a pair of Weyl cones. Specifically, we call a symmetric (normal) WSM if it is $P$ symmetric and an asymmetric WSM otherwise.

Next, we characterize the shape of a single Weyl cone by the discrete symmetries. We cannot use $C$ and $P$ symmetries separately because they flip chiralities. Relevant transformations are $T$ and $CP$, which are realized in the following way for a two-component Weyl fermion $\psi_{\bs{k}}$:
\begin{align}
 T\psi_{\bs{k}}T^{-1}&=-\sigma_2\psi_{-\bs{k}+\bs{k}_0},
 \\
 CP\psi_{\bs{k}}(CP)^{-1}&=\im\sigma_2\psi^*_{\bs{k}},
 \\
CPT\psi_{\bs{k}}(CPT)^{-1}&=-\im\psi^*_{-\bs{k}+\bs{k}_0}.
\end{align}
$T$ is an anti-unitary operator, which may be decomposed as $\tilde{T}\mathcal{K}$ with a unitary operator $\tilde{T}$ and an anti-unitary operator $\mathcal{K}$.
It is noted that $T$ is defined in such a way that the transformation is closed in a single Weyl cone by flipping $\bs{k}$ about the Weyl point $\bs{k}_0$ in each Weyl cone. We will set $\bs{k}_0=0$ in the following.

Now, we explore the transformation of the Weyl hamiltonian $H=\sum_{\bs{k}}\psi_{\bs{k}}^\dagger H_{\bs{k}}\psi_{\bs{k}}$ with $H_{\bs{k}}=\bs{k}\cdot\bs{\sigma}+g(\bs{k})$ in detail.
\begin{align}
 &THT^{-1}=\sum_{\bs{k}}\psi^\dagger_{-\bs{k}}\mathcal{K}(\sigma_2)^\dagger H_{\bs{k}}\sigma_2\mathcal{K}\psi_{-\bs{k}}.
\end{align}
Therefore, $T$-invariance $H=THT^{-1}$ requires 
\begin{align}
 &H_{\bs{k}}=\mathcal{K}\sigma_2 H_{-\bs{k}}\sigma_2\mathcal{K},
\end{align}
which leads to a condition for the Weyl hamiltonian
\begin{align}
 &\bs{k}\cdot\bs{\sigma}+g(\bs{k})
 =\bs{k}\cdot\bs{\sigma}+g(-\bs{k}).
\end{align}
Hence, $g(\bs{k})$ breaks $T$-symmetry if it is not an even function of $\bs{k}$.

$CP$-transformation of hamiltonian is
\begin{align}
 (CP)H(CP)^{-1}&=\sum_{\bs{k}}\psi^t_{\bs{k}}(\im\sigma_2)^\dagger H_{\bs{k}}\im\sigma_2\psi^*_{\bs{k}}
 \nonumber\\
 &=-\sum_{\bs{k}}\psi^\dagger_{\bs{k}}(\im\sigma_2)^t H^t_{\bs{k}}(\im\sigma_2)^*\psi_{\bs{k}}.
\end{align}
$H=(CP)H(CP)^{-1}$ requires
\begin{align}
 &H_{\bs{k}}=-(\im\sigma_2)^t H^t_{\bs{k}}\im\sigma_2,
\end{align}
which leads to
\begin{align}
 &\bs{k}\cdot\bs{\sigma}+g(\bs{k})
 =\bs{k}\cdot\bs{\sigma}-g(\bs{k}).
\end{align}
Any function $g(\bs{k})$ breaks $CP$-symmetry. This is the particle-hole symmetry for WSMs used in the main text.

Finally, we look at $CPT$-transformation.
$CPT$-transformation of hamiltonian is
\begin{align}
 (CPT)H(CPT)^{-1}&=-\sum_{\bs{k}}\psi^\dagger_{\bs{k}}\mathcal{K}H^t_{-\bs{k}}\mathcal{K}\psi_{\bs{k}}.
\end{align}
$H=(CPT)H(CPT)^{-1}$ requires
\begin{align}
 &H_{\bs{k}}=-\mathcal{K}H^t_{-\bs{k}}\mathcal{K},
\end{align}
which leads to
\begin{align}
 &\bs{k}\cdot\bs{\sigma}+g(\bs{k})
 =\bs{k}\cdot\bs{\sigma}-g(-\bs{k}).
\end{align}
$g(\bs{k})$ breaks $CPT$-symmetry if it is not an odd function of $\bs{k}$.

%
%



\bibliography{PVE}

\providecommand{\href}[2]{#2}\begingroup\raggedright\begin{thebibliography}{10}

\bibitem{Klinkhamer:2004hg}
F.~R. Klinkhamer and G.~E. Volovik, ``{Emergent CPT violation from the
  splitting of Fermi points},''
  \href{http://dx.doi.org/10.1142/S0217751X05020902}{{\em Int. J. Mod. Phys.}
  {\bfseries A20} (2005) 2795--2812},
\href{http://arxiv.org/abs/hep-th/0403037}{{\ttfamily arXiv:hep-th/0403037
  [hep-th]}}.

\bibitem{Volovik:2003fe}
G.~E. Volovik, ``{The Universe in a helium droplet},''
\href{http://dx.doi.org/https://global.oup.com/academic/product/the-universe-in-a-helium-droplet-9780198507826?cc=us&lang=en&}{{\em
  Int. Ser. Monogr. Phys.} {\bfseries 117} (2006) 1--526}.

\bibitem{Wan:2011udc}
X.~Wan, A.~M. Turner, A.~Vishwanath, and S.~Y. Savrasov, ``{Topological
  semimetal and Fermi-arc surface states in the electronic structure of
  pyrochlore iridates},''
  \href{http://dx.doi.org/10.1103/PhysRevB.83.205101}{{\em Phys. Rev.}
  {\bfseries B83} no.~20, (2011) 205101},
\href{http://arxiv.org/abs/1007.0016}{{\ttfamily arXiv:1007.0016
  [cond-mat.str-el]}}.

\bibitem{Fukushima:2008xe}
K.~Fukushima, D.~E. Kharzeev, and H.~J. Warringa, ``{The Chiral Magnetic
  Effect},'' \href{http://dx.doi.org/10.1103/PhysRevD.78.074033}{{\em Phys.
  Rev.} {\bfseries D78} (2008) 074033},
\href{http://arxiv.org/abs/0808.3382}{{\ttfamily arXiv:0808.3382 [hep-ph]}}.

\bibitem{Son:2012wh}
D.~T. Son and N.~Yamamoto, ``{Berry Curvature, Triangle Anomalies, and the
  Chiral Magnetic Effect in Fermi Liquids},''
  \href{http://dx.doi.org/10.1103/PhysRevLett.109.181602}{{\em Phys. Rev.
  Lett.} {\bfseries 109} (2012) 181602},
  \href{http://arxiv.org/abs/1203.2697}{{\ttfamily arXiv:1203.2697
  [cond-mat.mes-hall]}}.

\bibitem{Son:2012bg}
D.~T. Son and B.~Z. Spivak, ``{Chiral Anomaly and Classical Negative
  Magnetoresistance of Weyl Metals},''
  \href{http://dx.doi.org/10.1103/PhysRevB.88.104412}{{\em Phys. Rev.}
  {\bfseries B88} (2013) 104412},
  \href{http://arxiv.org/abs/1206.1627}{{\ttfamily arXiv:1206.1627
  [cond-mat.mes-hall]}}.

\bibitem{Zyuzin:2012tv}
A.~A. Zyuzin and A.~A. Burkov, ``{Topological response in Weyl semimetals and
  the chiral anomaly},''
  \href{http://dx.doi.org/10.1103/PhysRevB.86.115133}{{\em Phys. Rev.}
  {\bfseries B86} (2012) 115133},
  \href{http://arxiv.org/abs/1206.1868}{{\ttfamily arXiv:1206.1868
  [cond-mat.mes-hall]}}.

\bibitem{Basar:2013iaa}
G.~Basar, D.~E. Kharzeev, and H.-U. Yee, ``{Triangle anomaly in Weyl
  semimetals},'' \href{http://dx.doi.org/10.1103/PhysRevB.89.035142}{{\em Phys.
  Rev.} {\bfseries B89} no.~3, (2014) 035142},
\href{http://arxiv.org/abs/1305.6338}{{\ttfamily arXiv:1305.6338 [hep-th]}}.

\bibitem{vazifeh2013electromagnetic}
M.~M. Vazifeh and M.~Franz, ``{Electromagnetic Response of Weyl Semimetals},''
  \href{http://dx.doi.org/10.1103/PhysRevLett.111.027201}{{\em Phys. Rev.
  Lett.} {\bfseries 111} (2013) 027201},
  \href{http://arxiv.org/abs/1303.5784}{{\ttfamily arXiv:1303.5784
  [cond-mat.mes-hall]}}.

\bibitem{goswami2013axionic}
P.~Goswami and S.~Tewari, ``{Axionic field theory of $(3+1)$-dimensional Weyl
  semimetals},'' \href{http://dx.doi.org/10.1103/PhysRevB.88.245107}{{\em Phys.
  Rev. B} {\bfseries 88} (2013) 245107},
  \href{http://arxiv.org/abs/1210.6352}{{\ttfamily arXiv:1210.6352
  [cond-mat.mes-hall]}}.

\bibitem{Adler:1969gk}
S.~L. Adler, ``{Axial vector vertex in spinor electrodynamics},''
\href{http://dx.doi.org/10.1103/PhysRev.177.2426}{{\em Phys. Rev.} {\bfseries
  177} (1969) 2426--2438}.

\bibitem{Bell:1969ts}
J.~S. Bell and R.~Jackiw, ``{A PCAC puzzle: $\pi^0 \to \gamma\gamma$ in the
  sigma model},''
\href{http://dx.doi.org/10.1007/BF02823296}{{\em Nuovo Cim.} {\bfseries A60}
  (1969) 47--61}.

\bibitem{Nielsen:1983rb}
H.~B. Nielsen and M.~Ninomiya, ``{The Adler-Bell-Jackiw anomaly and Weyl
  fermions in a crystal},''
\href{http://dx.doi.org/10.1016/0370-2693(83)91529-0}{{\em Phys. Lett.}
  {\bfseries 130B} (1983) 389--396}.

\bibitem{Li:2014bha}
Q.~Li, D.~E. Kharzeev, C.~Zhang, Y.~Huang, I.~Pletikosic, A.~V. Fedorov, R.~D.
  Zhong, J.~A. Schneeloch, G.~D. Gu, and T.~Valla, ``{Observation of the chiral
  magnetic effect in ZrTe5},'' \href{http://dx.doi.org/10.1038/nphys3648}{{\em
  Nature Phys.} {\bfseries 12} (2016) 550--554},
\href{http://arxiv.org/abs/1412.6543}{{\ttfamily arXiv:1412.6543
  [cond-mat.str-el]}}.

\bibitem{kim2013dirac}
H.-J. Kim, K.-S. Kim, J.-F. Wang, M.~Sasaki, N.~Satoh, A.~Ohnishi, M.~Kitaura,
  M.~Yang, and L.~Li, ``{Dirac versus Weyl Fermions in Topological Insulators:
  Adler-Bell-Jackiw Anomaly in Transport Phenomena},''
  \href{http://dx.doi.org/10.1103/PhysRevLett.111.246603}{{\em Phys. Rev.
  Lett.} {\bfseries 111} (2013) 246603},
  \href{http://arxiv.org/abs/1307.6990}{{\ttfamily arXiv:1307.6990
  [cond-mat.str-el]}}.

\bibitem{xiong2015evidence}
J.~Xiong, S.~K. Kushwaha, T.~Liang, J.~W. Krizan, M.~Hirschberger, W.~Wang,
  R.~J. Cava, and N.~P. Ong, ``{Evidence for the chiral anomaly in the Dirac
  semimetal Na$_3$Bi},'' \href{http://dx.doi.org/10.1126/science.aac6089}{{\em
  Science} {\bfseries 350} no.~6259, (2015) 413--416}.

\bibitem{li2015giant}
C.-Z. Li, L.-X. Wang, H.~Liu, J.~Wang, Z.-M. Liao, and D.-P. Yu, ``{Giant
  negative magnetoresistance induced by the chiral anomaly in individual
  Cd$_3$As$_2$ nanowires},''
  \href{http://dx.doi.org/http://dx.doi.org/10.1038/ncomms10137}{{\em Nature
  Communications} {\bfseries 6} (2015) },
  \href{http://arxiv.org/abs/1504.07398}{{\ttfamily arXiv:1504.07398
  [cond-mat.mes-hall]}}.

\bibitem{huang2015observation}
X.~Huang, L.~Zhao, Y.~Long, P.~Wang, D.~Chen, Z.~Yang, H.~Liang, M.~Xue,
  H.~Weng, Z.~Fang, X.~Dai, and G.~Chen, ``{Observation of the
  Chiral-Anomaly-Induced Negative Magnetoresistance in 3D Weyl Semimetal
  TaAs},'' \href{http://dx.doi.org/10.1103/PhysRevX.5.031023}{{\em Phys. Rev.
  X} {\bfseries 5} (2015) 031023},
  \href{http://arxiv.org/abs/1503.01304}{{\ttfamily arXiv:1503.01304
  [cond-mat.mtrl-sci]}}.

\bibitem{wang2015helicity}
Z.~Wang, Y.~Zheng, Z.~Shen, Y.~Lu, H.~Fang, F.~Sheng, Y.~Zhou, X.~Yang, Y.~Li,
  C.~Feng, and Z.-A. Xu, ``{Helicity-protected ultrahigh mobility Weyl fermions
  in NbP},'' \href{http://dx.doi.org/10.1103/PhysRevB.93.121112}{{\em Phys.
  Rev. B} {\bfseries 93} (2016) 121112},
  \href{http://arxiv.org/abs/1506.00924}{{\ttfamily arXiv:1506.00924
  [cond-mat.mes-hall]}}.

\bibitem{zhang2015observation}
C.~Zhang, S.-Y. Xu, I.~Belopolski, Z.~Yuan, Z.~Lin, B.~Tong, N.~Alidoust, C.-C.
  Lee, S.-M. Huang, H.~Lin, {\em et~al.}, ``{Observation of the
  Adler-Bell-Jackiw chiral anomaly in a Weyl semimetal},''
\href{http://arxiv.org/abs/1503.02630}{{\ttfamily arXiv:1503.02630
  [cond-mat.mes-hall]}}.

\bibitem{yang2015observation}
X.~Yang, Y.~Li, Z.~Wang, Y.~Zhen, and Z.-a. Xu, ``{Observation of Negative
  Magnetoresistance and nontrivial $\pi$ Berrys phase in 3D Weyl semi-metal
  NbAs},''
\href{http://arxiv.org/abs/1506.02283}{{\ttfamily arXiv:1506.02283
  [cond-mat.str-el]}}.

\bibitem{shekhar2015large}
F.~Arnold, C.~Shekhar, S.-C. Wu, Y.~Sun, R.~D. dos Reis, N.~Kumar, M.~Naumann,
  M.~O. Ajeesh, M.~Schmidt, A.~G. Grushin, J.~H. Bardarson, M.~Baenitz,
  D.~Sokolov, H.~Borrmann, M.~Nicklas, C.~Felser, E.~Hassinger, and B.~Yan,
  ``{Negative magnetoresistance without well-defined chirality in the Weyl
  semimetal TaP},''
  \href{http://dx.doi.org/http://dx.doi.org/10.1038/ncomms11615}{{\em Nature
  Communications} {\bfseries 7} (2016) },
  \href{http://arxiv.org/abs/1506.06577}{{\ttfamily arXiv:1506.06577
  [cond-mat.mtrl-sci]}}.

\bibitem{yang2015chiral}
X.~Yang, Y.~Liu, Z.~Wang, Y.~Zheng, and Z.-a. Xu, ``{Chiral anomaly induced
  negative magnetoresistance in topological Weyl semimetal NbAs},''
\href{http://arxiv.org/abs/1506.03190}{{\ttfamily arXiv:1506.03190
  [cond-mat.mtrl-sci]}}.

\bibitem{Soluyanov:2015cn}
A.~A. Soluyanov, D.~Gresch, Z.~Wang, Q.~Wu, M.~Troyer, X.~Dai, and B.~A.
  Bernevig, ``{Type-II Weyl semimetals},''
  \href{http://dx.doi.org/http://dx.doi.org/10.1038/nature15768}{{\em Nature}
  {\bfseries 527} (2015) }, \href{http://arxiv.org/abs/1507.01603}{{\ttfamily
  arXiv:1507.01603 [cond-mat.mes-hall]}}.

\bibitem{Yong:2015}
Y.~Xu, F.~Zhang, and C.~Zhang, ``Structured weyl points in spin-orbit coupled
  fermionic superfluids,''
  \href{http://dx.doi.org/10.1103/PhysRevLett.115.265304}{{\em Phys. Rev.
  Lett.} {\bfseries 115} (Dec, 2015) 265304}.
  \url{https://link.aps.org/doi/10.1103/PhysRevLett.115.265304}.

\bibitem{fang2012multi}
C.~Fang, M.~J. Gilbert, X.~Dai, and B.~A. Bernevig, ``{Multi-Weyl Topological
  Semimetals Stabilized by Point Group Symmetry},''
  \href{http://dx.doi.org/10.1103/PhysRevLett.108.266802}{{\em Phys. Rev.
  Lett.} {\bfseries 108} (2012) 266802},
  \href{http://arxiv.org/abs/1111.7309}{{\ttfamily arXiv:1111.7309
  [cond-mat.mes-hall]}}.

\bibitem{kharzeev2016chiral}
D.~Kharzeev, Y.~Kikuchi, and R.~Meyer, ``{Chiral magnetic effect without
  chirality source in asymmetric Weyl semimetals},''
  \href{http://dx.doi.org/10.1140/epjb/e2018-80418-1}{{\em Eur. Phys. J.}
  {\bfseries B91} no.~5, (2018) 83},
\href{http://arxiv.org/abs/1610.08986}{{\ttfamily arXiv:1610.08986
  [cond-mat.mes-hall]}}.

\bibitem{ishizuka2016emergent}
H.~Ishizuka, T.~Hayata, M.~Ueda, and N.~Nagaosa, ``{Emergent Electromagnetic
  Induction and Adiabatic Charge Pumping in Noncentrosymmetric Weyl
  Semimetals},'' \href{http://dx.doi.org/10.1103/PhysRevLett.117.216601}{{\em
  Phys. Rev. Lett.} {\bfseries 117} (2016) 216601},
  \href{http://arxiv.org/abs/1607.06537}{{\ttfamily arXiv:1607.06537
  [cond-mat.mes-hall]}}.

\bibitem{chan2016photocurrents}
C.-K. Chan, N.~H. Lindner, G.~Refael, and P.~A. Lee, ``{Photocurrents in Weyl
  semimetals},'' \href{http://dx.doi.org/10.1103/PhysRevB.95.041104}{{\em Phys.
  Rev. B} {\bfseries 95} (2017) 041104},
  \href{http://arxiv.org/abs/1607.07839}{{\ttfamily arXiv:1607.07839
  [cond-mat.mes-hall]}}.

\bibitem{Mukhergee1:2018}
S.~P. Mukherjee and J.~P. Carbotte, ``Imaginary part of hall conductivity in a
  tilted doped weyl semimetal with both broken time-reversal and inversion
  symmetry,'' \href{http://dx.doi.org/10.1103/PhysRevB.97.035144}{{\em Phys.
  Rev. B} {\bfseries 97} (Jan, 2018) 035144}.
  \url{https://link.aps.org/doi/10.1103/PhysRevB.97.035144}.

\bibitem{Mukhergee2:2018}
S.~P. Mukherjee and J.~P. Carbotte, ``Doping and tilting on optics in
  noncentrosymmetric multi-weyl semimetals,''
  \href{http://dx.doi.org/10.1103/PhysRevB.97.045150}{{\em Phys. Rev. B}
  {\bfseries 97} (Jan, 2018) 045150}.
  \url{https://link.aps.org/doi/10.1103/PhysRevB.97.045150}.

\bibitem{Mukhergee3:2018}
S.~P. Mukherjee and J.~P. Carbotte, ``Anomalous dc hall response in
  noncentrosymmetric tilted weyl semimetals,'' {\em Journal of Physics:
  Condensed Matter} {\bfseries 30} no.~11, (2018) 115702.
  \url{http://stacks.iop.org/0953-8984/30/i=11/a=115702}.

\bibitem{Yamamoto:2015gzz}
N.~Yamamoto, ``{Chiral transport of neutrinos in supernovae: Neutrino-induced
  fluid helicity and helical plasma instability},''
  \href{http://dx.doi.org/10.1103/PhysRevD.93.065017}{{\em Phys. Rev.}
  {\bfseries D93} no.~6, (2016) 065017},
\href{http://arxiv.org/abs/1511.00933}{{\ttfamily arXiv:1511.00933
  [astro-ph.HE]}}.

\bibitem{Golub2017}
L.~E. Golub, E.~L. Ivchenko, and B.~Z. Spivak, ``Photocurrent in gyrotropic
  weyl semimetals,'' {\em JETP Letters} {\bfseries 105} no.~12, (Jun, 2017)
  782--785.

\bibitem{huang2016new}
S.-M. Huang, S.-Y. Xu, I.~Belopolski, C.-C. Lee, G.~Chang, T.-R. Chang,
  B.~Wang, N.~Alidoust, G.~Bian, M.~Neupane, D.~Sanchez, H.~Zheng, H.-T. Jeng,
  A.~Bansil, T.~Neupert, H.~Lin, and M.~Z. Hasan, ``{New type of Weyl semimetal
  with quadratic double Weyl fermions},''
  \href{http://dx.doi.org/10.1073/pnas.1514581113}{{\em Proceedings of the
  National Academy of Sciences} {\bfseries 113} no.~5, (2016) 1180--1185},
  \href{http://arxiv.org/abs/1503.05868}{{\ttfamily arXiv:1503.05868
  [cond-mat.mes-hall]}}.

\bibitem{Stephanov:2012ki}
M.~A. Stephanov and Y.~Yin, ``{Chiral Kinetic Theory},''
  \href{http://dx.doi.org/10.1103/PhysRevLett.109.162001}{{\em Phys. Rev.
  Lett.} {\bfseries 109} (2012) 162001},
  \href{http://arxiv.org/abs/1207.0747}{{\ttfamily arXiv:1207.0747 [hep-th]}}.

\bibitem{Chen:2012ca}
J.-W. Chen, S.~Pu, Q.~Wang, and X.-N. Wang, ``{Berry Curvature and
  Four-Dimensional Monopoles in the Relativistic Chiral Kinetic Equation},''
  \href{http://dx.doi.org/10.1103/PhysRevLett.110.262301}{{\em Phys. Rev.
  Lett.} {\bfseries 110} no.~26, (2013) 262301},
\href{http://arxiv.org/abs/1210.8312}{{\ttfamily arXiv:1210.8312 [hep-th]}}.

\end{thebibliography}\endgroup

\end{document}